\documentclass[twocolumn,prb]{revtex4}
\usepackage{graphicx}
\usepackage{amsmath}
\usepackage{amssymb}
\begin{document}

\title{Ultra-cold Neutron Production in Anti-ferromagnetic Oxygen Solid}

\author{C.-Y. Liu$^1$ and A. R. Young$^2$}

\affiliation{1. University of California, Los Alamos National Laboratory, Physics Division P-23, Los Alamos, NM 87545\\
2. North Carolina State University, Physics Department, Rayleigh, NC 27695 }

\date{\today}

\begin{abstract}

Spin waves, or magnons, in the anti-ferromagnetic $\alpha$ phase
of solid oxygen provide a novel mechanism for ultra-cold neutron (UCN) production.
Magnons dominate the energy exchange mechanisms for cold neutrons and UCN in 
solid $\alpha$-oxygen, much in the same way as do phonons in solid deuterium
superthermal UCN sources.
We present calculations of UCN
production and upscattering rates in S-O$_2$. 
The results indicate that S-O$_2$ is  
potentially a much more efficient UCN source material than solid deuterium.

\end{abstract}

\maketitle

\section{Introduction}

Ultra-cold neutrons (UCN) provide a low energy system in which to carry out 
fundamental neutron physics experiments to a much higher degree of 
precision than
that previously achieved with cold neutron experiments. 
UCN typically refer to neutrons with velocities less than 8 m/s.
Potential experiments using UCN include the measurements of neutron $\beta$-decay parameters 
to test the Standard Model,\cite{Gluck95,UCNA} the search for the neutron 
electric dipole moment to
test time reversal symmetry,\cite{Lamore,Ramsey} and the search for 
neutron--anti-neutron oscillations to
test baryon number conservation.\cite{Yuri} 
To date, some of these experiments have already been conducted with UCN but
most of them are still 
statistically limited. To increase the UCN flux,
Golub and Pendlebury\cite{Golub75} proposed a ``superthermal" 
source, in which the UCN production rate is enhanced through the efficient
inelastic coupling of cold neutrons to the phonon modes of a condensed matter
sample.
The UCN loss rate is suppressed in these sources by the reduction of the
thermal phonon population.  
Superthermal UCN sources have been successfully realized in 
super-fluid helium\cite{Ageron78} and solid deuterium.\cite{Morris02} 
Here, we propose solid oxygen (S-O$_2$) as a superthermal source, which 
shows considerable promise for enhanced UCN production via a novel mechanism. 

Solid $\alpha$-$^{16}$O$_{2}$ has 
manifest magnetic properties. 
Its ability to interact with cold neutrons and UCN, 
through neutron-magnon inelastic scattering, provides potential 
advantages which stem from the energy dispersion spectrum 
of its anti-ferromagnetic magnons.  
The large neutron-magnon inelastic scattering rate
together with oxygen's small nuclear absorption cross section suggest
that a S-O$_2$ UCN source may well compete favorably with 
solid deuterium UCN sources. 

This report is organized as follows. In section II, we calculate the 
neutron-magnon and neutron-phonon scattering cross sections in solid
oxygen, after reviewing the relevant magnetic properties and magnon
dispersion relations. In section III we compare the predicted performance
of a solid oxygen UCN source with the known performance of typical 
solid deuterium sources, and discuss possible mechanisms for UCN loss. 
We summarize our results in section IV.

\section{Neutron-Magnon Scattering}

\subsection{Magnetic Scattering Length}

In solid oxygen at low temperatures (T$<$23.9K), thermal
fluctuations quench and
long range anti-ferromagnetic ordering is established. 
This results from a slight distortion of the lattice structure
from a rhombohedral ($\beta$ phase) to a monoclinic symmetry ($\alpha$ phase).
The resulting 2-D anti-ferromagnetic
structure exhibits spin wave excitations.\cite{solid_oxygen2,solid_oxygen}
For our work on UCN, we will be concerned with S-O$_2$ in the $\alpha$-phase below 10K, where superthermal UCN production is possible.

The magnetism of solid oxygen originates from the unpaired $2p$ electrons
in the oxygen atom, two of which align and form a molecular ground state
$^{3}\Sigma ^{-}_{g}$ with a net spin $S=1$ and an orbital angular momentum $l=0$.
For a diatomic molecule made of two
$^{16}$O atoms (naturally abundant), the null nuclear spin and antisymmetric electronic wave-function
lead to a selection rule allowing only rotational states with odd 
parity.\cite{raman}
The absence of half of the rotational states is an
advantage in the application of S-O$_2$ as a UCN source, as will be discussed later.

A neutron interacts magnetically with an oxygen molecule. The   
interaction potential can be described as 
\begin{equation}
V = \mu \cdot H = \mu \cdot \nabla \left( \frac{\mu_e \times R}{|R|^3} \right),
\label{potential}
\end{equation}
\noindent where  $\mathbf{\mu} = \gamma \mu_N \sigma$ is the neutron magnetic 
moment with neutron spin $\sigma$, and  $\mu_e = -2 \mu_B \hat{s}$
is the electron magnetic moment with electron spin $\hat{s}$.
This magnetic coupling, which is the origin of 
neutron-magnon scattering, has an intrinsic scattering length
of 5.4 fm.\cite{Lovesey84} This is comparable to typical 
neutron-nuclear scattering lengths, involved in neutron-phonon scatterings. 

The differential cross section for neutron scattering 
in the presence of the potential (\ref{potential}) can be written:  
\begin{eqnarray}
\frac{d^2\sigma}{d\Omega _f dE} &=& \frac{k_f}{k_i} |\langle k_f 
\lambda _f |V|k_i \lambda _i \rangle|^2
 \delta(\hbar \omega + E_{\lambda'}-E_{\lambda}) \nonumber \\
\ &=& (\frac{m_n}{2\pi \hbar^2})^2 (2\gamma \mu_N \mu_B)^2 (4\pi)^2 \frac{k_f}{k_i} 
\sum _{\lambda _i\lambda _f \sigma _i \sigma _f} p_{\lambda _i}p_{\sigma _i} \nonumber \\
\ & & \times |\langle \lambda _i \sigma_i|\hat{\sigma _i} \cdot \hat{Q}|\lambda _f \sigma _f\rangle|^2
\delta(\hbar \omega + \Delta E) \nonumber \\
\ &=& r^2_0 \frac{k_f}{k_i} \mathbb{S}(\vec{\kappa},\omega).
\label{double_diff}
\end{eqnarray}
\noindent Here $\vec{k_i}$ is the initial neutron momentum vector, oriented
relative to a crystal axis by a solid angle $\Omega _i$, 
$\Omega _f$ is the solid angle for final neutron momentum $k _f$ relative to 
the same crystal axis, and $\lambda _i$, $\lambda _f$
are the initial and final states of the oxygen molecule. The above formula 
averages over  
the initial states of the target molecule and the neutron spin. 
From Eq.(\ref{potential}), the interaction matrix element is reduced to 
a scalar product between the neutron spin and the coherent 
lattice spin form factor $\hat{Q}$, which is defined using the   
neutron momentum transfer $\vec{\kappa}=\vec{k}_f-\vec{k}_i$ 
and the electron spin $\hat{s}_i$ as 
\begin{equation}
\hat{Q} = \sum_{i} e^{i \kappa \cdot r_i} \tilde{\kappa} \times (\hat{s_i} \times \tilde{\kappa}).
\end{equation}
The differential cross-section is conventionally simplified as 
a product of the square of the scattering length, the
kinematic factor $\frac{k _f}{k _i}$ and the scattering law $\mathbb{S}(\kappa,\omega)$.\footnote{The scattering law S($\kappa$,$\omega$) defined here is slightly
different from that defined in Ref~\cite{solid_oxygen} which only retains the dynamical
structure factor.} 
The effective scattering length $r_0$, characterizing the strength
of magnetic neutron scattering, is given by 
%\begin{eqnarray*}
%\ r_0 &=& \frac{m_n}{2\pi \hbar^2} \cdot 2 \gamma \mu_N \mu_B \cdot 4\pi \\
%\ &=& \frac{m_n}{2\pi \hbar^2} \cdot 2 \gamma \frac{e \hbar}{2 m_p c} \frac{e
%\hbar}{2 m_e c} \cdot 4\pi  = \gamma \frac{e^2}{m_e c^2},
%\end{eqnarray*}
\begin{equation}
r_0 = \frac{m_n}{2\pi \hbar^2} \cdot 2 \gamma \mu_N \mu_B \cdot 4\pi = \gamma \frac{e^2}{m_e c^2},
\end{equation}
\noindent where $\frac{e^2}{m_e c^2} =  2.82 fm $ is the 
the classical electron radius and $\gamma$=~-1.91 is  
the neutron gyro-magnetic ratio.
The detailed physics of neutron interactions in solid O$_2$ is absorbed in the 
scattering law $\mathbb{S}(\kappa,\omega)$.   

\subsection{Anti-ferromagnetic Magnons in S-O$_2$}

\begin{figure}[t]
\begin{center}
\includegraphics[width=3 in]{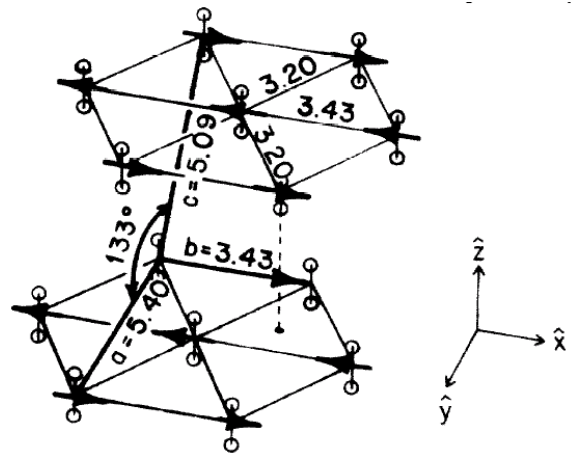}
\end{center}
\caption{A schematic of the lattice structure of solid $\alpha$-O$_2$.\cite{solid_oxygen}}
\label{figu_lattice}
\end{figure}

Calculation of the scattering law requires knowledge of the 
energy dispersion relation in the 
magnetic lattice of S-O$_2$.
As shown in Fig.\ref{figu_lattice}, solid $\alpha$-O$_{2}$ has a primitive unit cell with 
a distorted hexagon geometry.\cite{solid_oxygen}
There are 14 oxygen molecules in the unit
cell. On the basal plane, the b-axis is defined to be along the magnetization direction (which is also along one side of the hexagon), 
and the a-axis is assigned to be orthogonal to the b-axis. 
Due to a larger lattice constant along the c-axis,   
couplings between magnetization planes
were found to be negligible compared to the in-plane couplings. 
The dominant 
2-D magnetic structure are comprised of two interlacing
sub-lattices, each carrying the opposite
magnetization. Every lattice site is surrounded by 4 nearest neighbors on 
the opposite magnetic sub-lattice, and 2 next-nearest neighbors
on the same magnetic sub-lattice.

This 2-D anti-ferromagnetic system can be described by the Heisenberg hamiltonian  
\begin{equation}
H = -2 \sum_{<ij>} J_{ij} s_i \cdot s_j + \sum_{i}(-Ds^2_{xi}-D's^2_{yi}+D's^2_{zi}),
\label{Spin_H}
\end{equation}
\noindent where $J_{ij}$ characterizes the strength of exchange interactions
between molecular spin s$_i$ on site $i$ and s$_j$ on site $j$. 
$D$ represents the self-energy of the longitudinal spin,  
and $D'$ the self-energy of the transverse spin (here, the b- and a-axes are
parallel to the x and y-axes, respectively).
Note that even though the out-of-plane
coupling is weak, the self-energy term $D's_{z}^2$ alone in the hamiltonian
can result in significant contributions to scatterings in the z-direction. 

The hamiltonian of Eq.(\ref{Spin_H}) leads to
the dispersion relation of magnon excitations\cite{solid_oxygen}:
\begin{equation}
\omega_a(\vec{q}) 
= \sqrt{(2J_0 - 2J_{\vec{q}} -2J'_0 +D)^2- [2J'_{\vec{q}}+(-1)^a D']^2}.
\label{disp}
\end{equation}
\noindent Here $\omega_a(\vec{q}) \equiv \omega_{q,a}$ is the energy of a 
magnon, with momentum $q$.
The angular dependence of the dispersion relation 
comes from $J_{\vec{q}}$ and $J'_{\vec{q}}$, which are the Fourier transform 
of the coupling coefficients $J_{ij}$: $J_{\vec{q}}=\sum_j
J_j e^{i\vec{q} \cdot \vec{r}_i}$ defined on the sites on one sub-lattice,
and $J'$ defined in the same way but on the opposite sub-lattice. 
Following the analysis of Stephens and Majkrzak,\cite{solid_oxygen} we take 
the numerical values of
the spin coupling $J_{NN}$ between 
the nearest neighbors to be -2.44~meV, the value $J_{NNN}$ between the next 
nearest neighbors to be -1.22~meV, and $J_{\perp}$ for the out-of-plane 
(z direction) coupling to be zero. 
Also, we use $D$=0.132 meV and $D'$=0.118 meV.
Here, $a=\{0,1\}$ is the branch index of magnon excitations, which include
one acoustic and one optical mode. 

\begin{figure}[t]
\begin{center}
\includegraphics[width=3.4 in]{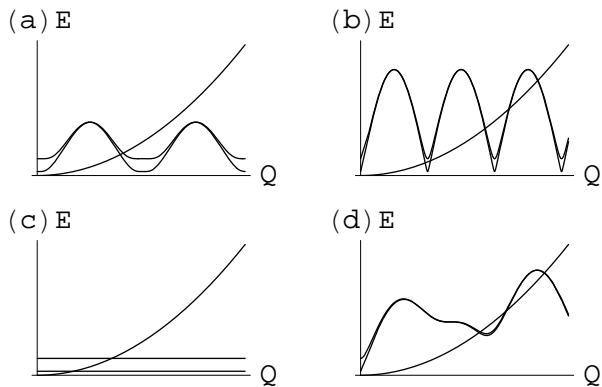}
\end{center}
\caption{The dispersion relation (Eq.(\ref{disp})) of magnons in solid $\alpha$-O$_2$
plotted for momentum transfer along (1,0,0) in (a), (0,1,0) in (b), (0,0,1) in (c), and (1,1,1) in (d).
There are two branches (acoustic and optical).
In (a,b,c) different periodicities originate from different lattice constants 
on these crystal axes. In (d), momentum transfers along a non-symmetry 
axis produce a non-periodic
dispersion relation. Intersections with the parabolic
free neutron dispersion set the four momentum (energy, momentum) 
for UCN production.}
\label{figu_DRmagnon}
\end{figure}

Fig.~\ref{figu_DRmagnon} illustrates the dispersion relation
at several different neutron scattering angles, together with   
the free neutron energy $E_n=\hbar^2\kappa ^2/2m_n$. 
For certain values of the incident neutron momentum, incident neutrons can
lose essentially all their kinetic energy in the excitation of single 
magnon, and emerge as (or ``downscatter" to) UCN. These values are determined
by the points where  
the free neutron curve intersects 
the magnon dispersion curves in Fig.~\ref{figu_DRmagnon}. 
Note that this process is coherent, involving the  
simultaneous conservations of energy and momentum in the scattering
process, and is analogous to the creation of UCN via phonon production
in superfluid $^4$He. 
We are only interested
in the scattering conditions in which UCN are produced in the final state.
Processes involving the creation of multiple magnons are also possible, 
but these are higher order effects which we expect to result in corrections 
on the level
of a few percent and we neglect them in our calculation.

The energies of typical excitations in condensed matter systems range from
tens to hundreds of Kelvin. This is comparable to the kinetic energy 
of cold neutrons. Cold neutrons are therefore used in UCN production 
experiments as they can efficiently
couple to the condensed matter excitation modes and exploit the available 
energy exchange channels. Our goal is to transform as many 
incident cold neutrons into UCN as possible. Neutron downscattering through 
excitations in a bulk material is known to be the most efficient mechanism
by far for this process. 

In a solid target, the 
anisotropy of the lattice structure manifests
itself in an orientation-dependent dispersion relation, as illustrated in 
Fig.\ref{figu_DRmagnon}.
As a result, for every neutron momentum transfer $\kappa$, 
there exists a continuum of energy modes available for excitation.
A large fraction of the incident cold neutron spectrum
can therefore be used for UCN production.
This is a potential advantage to the case of
neutron scattering in super-fluid $^4$He (the first proposed UCN source), 
where only a narrow band of 
incident neutrons with energy centered at 11 K can produce UCN. This 
results from the degenerate dispersion curves of phonon 
excitations in the super-fluid, which is an isotropic media.

\begin{figure}
\begin{center}
\includegraphics[width=3.3 in]{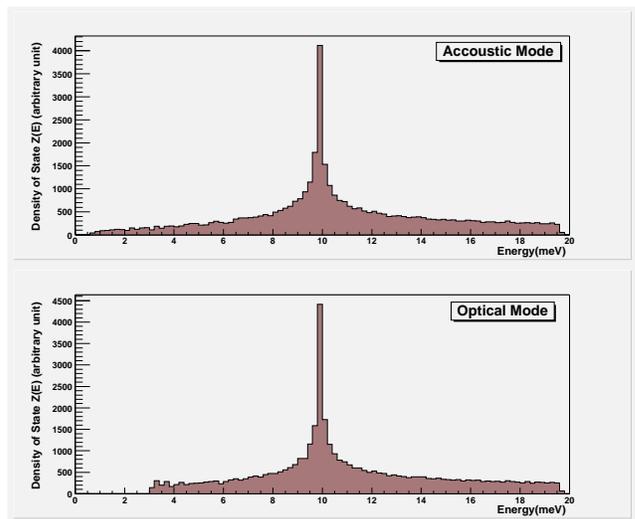}
\end{center}
\caption{The density of states of magnon excitation modes in
solid $\alpha$-O$_2$, dark histogram for the acoustic mode and light for the
optical mode.} 
\label{figu_ZSO2}
\end{figure}

When sampled
uniformly over the 3-D magnon momentum space $\vec{q}$,  
the density of magnon states $Z(E)$ can be histogramed over the energy scale,
according to the allowed energy given by the dispersion 
relation Eq.(\ref{disp}), 
i.e., 
\begin{equation}
\sum_{\vec{q}}  = \int dE Z(E).
\end{equation}
\noindent The density of magnon states in S-O$_2$ is plotted in Fig.~\ref{figu_ZSO2}. 
A distinct peak at 10 meV results from the  
flattening of the dispersion curve approaching the Brillouin zone (BZ) 
boundaries of 
$\kappa _x$. One would also expect a 20 meV peak 
corresponding to the BZ boundary of $\kappa _y$, however, this peak  
is less significant because of the smaller
flattening region due to the higher periodicity of $\kappa _y$. 
Note that 
there are no magnon modes with energies below E$_{0}$ = 0.7 meV.
The absence of low energy magnon modes 
is particularly important for UCN upscattering events,  
and we will return to this point when discussing UCN loss
in Sec. III.

\subsection{Scattering Law in AFM crystals}

The reference by Lovesey,\cite{Lovesey84} contains detailed calculations
liking the scattering matrix elements to the scattering law
in Eq.(\ref{double_diff}).  
The scattering law for neutorn inelastic scattering from 
a Heisenberg anti-ferromagnet with single magnon creation(+) and annihilation($-$) is given as
\begin{eqnarray}
\lefteqn{ \mathbb{S}^{\pm}(\vec{\kappa},\omega) = 
[ \frac{1}{2} g F(\kappa) ]^2 \frac{1}{4} (1+\tilde{k_x}^2) e^{-2W(\vec{\kappa})} 
\frac{(2\pi)^3}{Nv_0} } \nonumber \\
&& \sum_{a=0,1}\sum_{q,\tau} (n+\frac{1}{2}\pm \frac{1}{2})
\delta (\hbar \omega_{q,a} \mp \hbar \omega)\delta^{3}(\vec{q} \mp \vec{\kappa} - \vec{\tau})  \nonumber \\
&& \qquad \qquad \qquad \times 
\{ u^2_{q}+v^2_{q}+2 u_{q} \cdot v_{q} \cos( \vec{\rho} \cdot \vec{\tau}) \},
\label{magnon_diff}
\end{eqnarray}
\noindent where $F(\kappa)$ is the electron spin form factor of an oxygen 
molecule, $g=2$ is the electron gyro-magnetic ratio, $\tilde{k}_{x}$ is the unit neutron 
momentum vector projection on the magnetization direction, $N$ is the total number of
scatterers, $v_0$ is the unit volume occupied per scatterer, and $n$
is the magnon occupation number. For the zeroth order
approximation,\cite{Kleiner55} we use the spherical
Bessel function\cite{Temme} of order zero, $j_{0}(\kappa R/2)$, 
for the electron spin form factor $F(\kappa)$. The internuclear distance $R$ is
1.21 $\AA$.
\footnote{Although difficult to quantify, we
believe that our calculations are accurate within a factor of 2.}
The additional terms $u_q$ and $v_q$ are due to the two sets of spin 
waves in the opposite sublattices, and they can be 
reduced to 
\begin{eqnarray}
\lefteqn{ \{ u_{q} + (-1)^m v_{q} \}^2 }  \\
&=& 2sN \frac{2J_0-2J'_0-2J_{q}+D-(-1)^m[2J'_{q}+(-1)^aD']}{E_a(q)}, \nonumber
\end{eqnarray}
\noindent with $m=0$ for a nuclear peak, and $m=1$ for a magnetic peak.

Summing over the allowed momentum transfer $q$, Eq.(\ref{magnon_diff})
can be further simplified to terms involving a 
discrete sum over allowed reciprocal lattice momentum $\tau$: 
\begin{eqnarray}
\label{eq_momentum}
\lefteqn{ \frac{(2\pi)^3}{Nv_0} \sum_{q,\tau} \delta^3(\kappa+q-\tau)} \\
&=& \frac{(2\pi)^3}{Nv_0} \sum_{\tau}
\frac{V}{(2\pi)^3} \int d^3q \delta^3(\kappa+q-\tau) 
= \sum_{\tau} \delta_{\tau,\kappa+q}.\nonumber
\end{eqnarray}

\subsection{Evaluation of Neutron Coherent Scattering from Polycrystal}

\begin{figure}[t]
\begin{center}
\includegraphics[width=3.3 in]{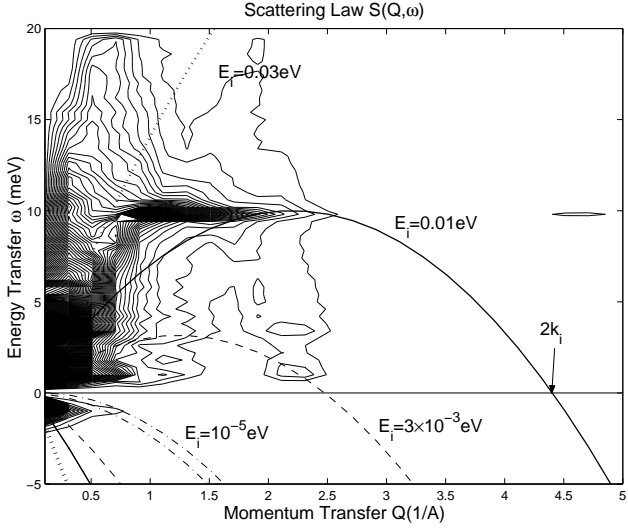}
\end{center}
\caption{A contour plot of the neutron scattering law $S(\kappa,\omega)$
in a 5~K solid $\alpha$-O$_{2}$:
The upper half plane($\omega>0 \rightarrow E_i > E_f$) covers
neutron downscatterings, and the $\omega<0$ covers neutron upscatterings.
The area of integration is bounded by the dotted lines for
incident neutrons with an initial energy E$_i$=0.03~eV, and solid lines, dashed
lines and dash-dotted lines, for E$_i$=0.01~eV, 3$\times$10$^{-3}$~eV and 10$^{-5}$~eV,
respectively.}
\label{figu_contourSS}
\end{figure}

We are now ready to perform the cross-section calculations with the 
scattering law $\mathbb{S}(\kappa,\omega)$ for S-O$_2$. 
We present here our algorithm for  
calculating the pure coherent scattering cross-sections.
We are interested only in the total cross-section which, according to Eq.(\ref{double_diff}), is
\begin{equation}
\sigma^{tot}(E_{i},\Omega_i) = r_{0}^{2} \int_{0}^{\infty} dE_{f} \int d\Omega_f 
\frac{k_f}{k_i} \mathbb{S}(\vec{\kappa},\omega). 
\end{equation}
\noindent To consolidate the number of variables,
we apply a coordinate transformation ($\Omega$, $E_f$) $\rightarrow$ ($\vec{\kappa}$, $\omega$) =
($\phi$, $\kappa$, $\omega$) with
\begin{eqnarray*}
&\mbox{}& \left\{ \begin{array}{lll}
\omega &=& E_i-E_f \\
 \kappa &=& \sqrt{k_f^2+k_i^2-2k_f k_i cos\theta}=  
\end{array} \right.  \\
&\mbox{}& \hspace{1cm} \sqrt{\frac{2m}{\hbar^2}}\sqrt{E_f+E_i-2\sqrt{E_f E_i} \mu}
\end{eqnarray*}
\noindent This leads to
\begin{equation}
d\Omega dE_f = (d\phi d\mu )dE_f = d\phi \Big( \frac{\kappa}{k_f k_i} 
d\omega d\kappa \Big).
\end{equation}
\noindent Here, the neutron energy transfer $\omega$ is also the magnon
energy gain, and $\kappa$ represents the magnitude of neutron momentum transfer.
The total scattering cross-section becomes 
\begin{equation}
\sigma^{tot}(E_i,\Omega_i) = r_{0}^2 \int\!\!\!\int\!\!\!\int d\omega d\kappa d\phi \frac{\kappa}
{k_f k_i} \frac{k_f}{k_i} \mathbb{S}(\vec{\kappa},\omega),  
\label{integral1}
\end{equation}
\noindent which can be further reduced to
\begin{equation}
\sigma^{tot}(E_i) = 
2\pi  r_{0}^2 \frac{\hbar^2}{2mE_i} \!\int_{-E_{BZ}}^{min(E_i,E_{BZ})}\!\!\! 
d\omega \!\!\!\int_{\kappa _{lower}}^{\kappa _{upper}}\!\!\! d\kappa  \kappa 
\mathcal{S}(\kappa,\omega). 
\label{integral}
\end{equation}
\noindent In a typical material with a periodic lattice structure, 
the dispersion relation contains the  
maximum energy at the Brillouin zone boundary, i.e., $\omega_{max}=E_{BZ}$.
In the case of single magnon exchange, this
defines the lower bound (energy gain) of the energy transfer integral. 
The upper bound (energy loss) can be up to $E_{BZ}$ or the initial 
energy of the neutron. 
The lower, $\kappa_{lower}$, and upper, $\kappa_{upper}$, bounds in the 
momentum integral correspond to forward and backward scattering events,
respectively:
for a given magnitude of energy transfer and initial neutron energy,
\begin{eqnarray}
\kappa^2 \geq \frac{2m}{\hbar^2} (E_f+E_i-2 \sqrt{E_f E_i}) \equiv \kappa^2_{lower} \nonumber \\
\kappa^2 \leq \frac{2m}{\hbar^2} (E_f+E_i+2 \sqrt{E_f E_i}) \equiv \kappa^2_{upper}. \end{eqnarray}
\noindent The bounds
\begin{eqnarray}
|\kappa_{lower}| = \sqrt{\frac{2m}{\hbar^2}} \left| \sqrt{E_i-\omega}-\sqrt{E_i} \right| \\
|\kappa_{upper}| = \sqrt{\frac{2m}{\hbar^2}} (\sqrt{E_i-\omega}+\sqrt{E_i} )
\end{eqnarray}
\noindent are plotted, together with a contour plot of the scattering law, 
in Fig.\ref{figu_contourSS}. The area bounded by
these two curves defines the integration area.

Note that in the reduction Eq.(\ref{integral1}) to Eq.(\ref{integral}),
we have averaged the scattering law $\mathbb{S}(\vec{\kappa},\omega)$
over the incident neutron angle $\Omega_i$, i.e.,
\begin{equation}
\ \mathcal{S}(\kappa ,\omega) = \frac{1}{4\pi} \int d\Omega_i \mathbb{S}(\vec{\kappa},\omega).
\end{equation}
\noindent The reduced scattering law,
$\mathcal{S}(\kappa,\omega)$, is made isotropic, so that
the double scalar variable integration in Eq.(\ref{integral}) becomes
straightforward. Note that
even though it results in the same form for the final integral,
the treatment presented here does not make use of  
the ``incoherent approximation'' widely used in other cross-section calculation packages.\cite{LEAPR}
The incoherent approximation relaxes the momentum conservation essential
to coherent processes, and therefore it is subject to errors in the low energy 
limit.
Here, we arrive at Eq.(\ref{integral}) after integrating the 
scattering law $\mathbb{S}(\vec{\kappa},\omega)$ over the incident angle, 
and this is 
suitable for the case of neutron scattering off polycrystals.

Eq.(\ref{integral}) is evaluated numerically using a Monte-Carlo technique.
We construct a table of values for
the scattering law $\mathcal{S}(\kappa ,\omega)$ using the  
dispersion relation $\omega(\vec{\kappa})$ given in Eq.(\ref{disp}).
With a given $k$, the vector
$\vec{\kappa}$=($\kappa \sin{\theta}\cos{\phi}$, $\kappa \sin{\theta}\sin{\phi}$,
$\kappa \cos{\theta}$) is specified for a random choice of orientation angles
($\theta$,$\phi$);
the dispersion relation $\omega(\vec{\kappa})$ gives the corresponding 
energy of excitation, $\omega$.  
Momentum transfers
larger than the BZ momentum are mapped back to the first BZ, 
according to  Eq.(\ref{eq_momentum}), via unique Bloch translations. 
A numerical value of $\mathcal{S}(\kappa _i,\omega_j)$ is then calculated
and added to a running total for each bin $(i,j)$ in the parameter space. 
This process is repeated, stepping through the parameter space according to 
$i\Delta \kappa = \kappa _i$ and $j\Delta\omega \leq \omega \leq (j+1)\Delta\omega$. 
Finally, we implement the angular average by repeating the Monte Carlo 
random angle assignment for each $\kappa$, and histogram the corresponding $\mathcal{S}(\kappa,\omega)$
values.

Combining all the ideas above,
we proceed to integrate Eq.(\ref{integral}) numerically
by summing over all kinematically accessible points on a grid in the
($\omega$, $q$) phase space. In the numerical approximation, 
the neutron downscattering cross-section is given by  
\begin{eqnarray}
\sigma^{tot}_{down}(E_i) &\cong& 2\pi r_0^2 \frac{\hbar^2}{2mE_i} \sum_{l(\omega_l>0)}
^{l(\omega_l=min(E_i,E_{BZ}))} \Delta \omega_{i} \nonumber \\
 &\times& \sum_{m(\kappa _m=\kappa _{lower})}
^{m(\kappa _m=\kappa _{upper})} \Delta \kappa _{m} \kappa _{m}
\mathcal{S}(\kappa _m,\omega_l)  
\label{discrete_down}
\end{eqnarray}
\noindent and the upscattering cross-section is 
\begin{eqnarray}
\sigma^{tot}_{up}(E_i) &\cong& 2\pi r_0^2 \frac{\hbar^2}{2mE_i} \sum_{l(\omega_l>0)}
^{l(\omega_l=E_{BZ})} \Delta \omega_{i} \nonumber \\
 &\times& \sum_{m(\kappa _m=\kappa _{lower})}
^{m(\kappa _m=\kappa _{upper})} \Delta \kappa _{m} \kappa _{m}
\mathcal{S}(\kappa _m,-\omega_l).
\label{discrete_up}
\end{eqnarray}

\begin{figure}[t]
\begin{center}
\includegraphics[width=3.3 in]{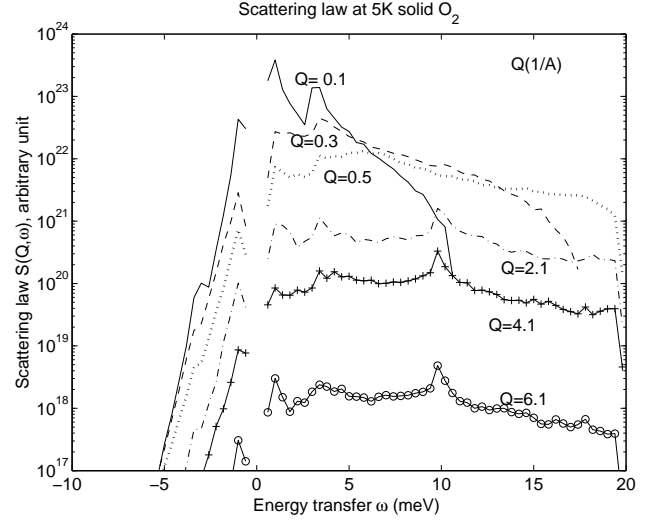}
\end{center}
\caption{Neutron scattering law $\mathcal{S}(\kappa,\omega)$ in a 5~K solid $\alpha$-O$_2$:
$\mathcal{S}(\kappa,\omega)$
is plotted as a function of energy transfer $\omega$ for momentum transfer $Q$=0.1,
0.3, 0.5, 2.1, 4.1 and 6.1 $\mbox{\AA}^{-1}$. Distinct energy dependences are shown for $\omega>0$
downscattering and $\omega<0$ upscattering parts.} 
\label{figu_SS_detail}
\end{figure}

For the results presented here, we constructed a table of 
$\mathcal{S}(\kappa,\omega)$ values
on a grid of 500 $\times$ 500 points, equally spaced by 
$\Delta \kappa=0.02\mbox{\AA}$ and $\Delta\omega=0.04$ meV.
The scattering law in a 5~K sample of solid oxygen is presented as a 
contour plot in 
Fig.~\ref{figu_contourSS}, and the detailed dependence of $\mathcal{S}(\kappa,\omega)$ on energy and momentum
transfer is illustrated in Fig.~\ref{figu_SS_detail}. 
Notice that in both figures, at low $\kappa$, there exist two
dominant contributions at 0.7~meV and 3~meV, which correspond to the 
self-energy terms involving $D$ and $D'$
of the acoustic and optical modes, respectively. As $\kappa$ increases, the
peak in the scattering law shifts to a higher value 
centered at 10~meV, which is clearly
the BZ boundary energy, as discussed before.

\begin{figure}[t]
\begin{center}
\includegraphics[width=3.3 in]{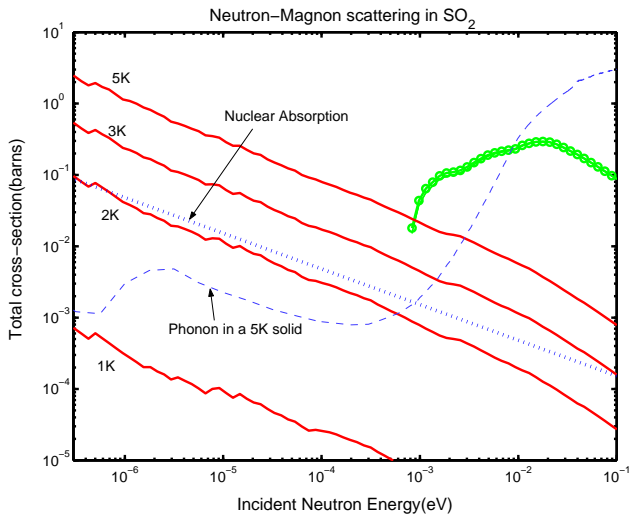}
\end{center}
\caption{Cross sections as a function of the incident neutron energy:
Solid curves: neutron upscattering from solid
$\alpha$-O$_2$ at temperatures of 1~K, 2~K, 3~K and 5~K with annihilation
of a single magnon. Circled curves: neutron downscattering via creation of 
one magnon from solid oxygen of 1~K, 2~K, 3~K and 5~K. Note that the downscattering
cross section is nearly independent of the solid temperature.
Dashed curve: neutron-phonon scattering in solid oxygen. 
Dotted curve: nuclear absorption cross-section per $^{16}$O$_{2}$ molecule.} 
\label{figu_total}
\end{figure}

Double integration of $\mathcal{S}(\kappa,\omega)$ using Eq.(\ref{discrete_down})
and Eq.(\ref{discrete_up}) then produce the downscattering and upscattering
cross sections for neutrons with a given initial energy. The sum of the two
gives the total cross sections. 
The resulting neutron-magnon scattering total cross sections in solid
oxygen at various temperatures below E$_0$ are presented in 
Fig.~\ref{figu_total}. 

\subsection{Phonon Scattering in S-O$_2$}

Neutron-phonon inelastic scattering is also present inside
solid $\alpha$-O$_2$.  
The scattering law for 
single phonon creation(+) and annihilation($-$) is\cite{Lovesey84}
\begin{eqnarray}
\label{phonon_diff}
\lefteqn{ \mathbb{S}^{\pm}(\vec{\kappa},\omega) = 
\frac{1}{2M_{s}} e^{-2W(\vec{\kappa})} 
\frac{(2\pi)^3}{v_0} \sum_{j=1,2,3} } \\
&& \sum_{q,\tau} \frac{|\vec{\kappa} \cdot \sigma^j(q)|^2}{\omega_{j}(q)} 
(n+\frac{1}{2} \pm \frac{1}{2})
\delta (\hbar \omega_{q,a} \mp \hbar \omega) \delta^{3}(\vec{q} \mp \vec{\kappa} - \vec{\tau}). \nonumber
\end{eqnarray}
%\noindent  The size of a practical UCN source is
%such that we do not expect more than
%one phonon or magnon to be created during the residency time of a neutron
%inside the converter, so we do not concern ourselves with multiple
%phonons, multiple magnons, or phonon-magnon events.

The neutron-phonon scattering cross-sections are
calculated using LEAPR, one of the subroutines in the NJOY package developed in
the Los
Alamos National Laboratory.~\cite{LEAPR}  The incoherent approximation is
used for the oxygen nucleus O$^{16}$, even though it is a pure 
coherent scatterer. While we still expect a discrepancy at low 
energies, we expect the incoherent approximation to yield results 
differing from the actual phonon cross section by at most a small 
numerical factor\cite{thesis}.
An iterative expansion is used to include creation and
annihilation of up to 100 phonons.
The cross section for neutron-phonon scattering is shown along with
those for neutron-magnon scattering in Fig.\ref{figu_total}.

Note that the above formula is similar to Eq.(\ref{magnon_diff}), 
with the important distinction of an additional inverse dependence on the scatterer mass
$M_{s}$. This indicates that phonon production is suppressed due to the 
smaller amplitudes
associated with heavier nuclear recoil. Conversely, magnon excitation  
takes place via neutron-electron interactions, and thus does not 
have the mass suppression. To be more precise, we will compare the prefactors
and the energy terms
in Eq.(\ref{magnon_diff}) and Eq.(\ref{phonon_diff}) with the corresponding 
scattering length: for the magnon part, one finds
\begin{eqnarray}
r_0^2 [\frac{1}{2} g F(\kappa)]^2 \frac{1}{4} (1+\tilde{k_x}^2) \frac{O(4J_{NN})}{\omega_{a}(q)} \nonumber \\
\sim r_0^2 \times O(1) \times \frac{1}{3} \times \frac{10 meV}{10 meV};
\nonumber
\end{eqnarray}
\noindent for the phonon part, one has to pay attention to the detail
that there are in fact two nuclei in one molecular scatterer, which 
contribute to the scattering amplitude coherently. Therefore, the
relevant term is
\begin{eqnarray}
(b+b) ^2 \frac{1}{2M_{s}}  \frac{|\vec{\kappa} \cdot \sigma^j(q)|^2}{\omega_{j}(q)}
&\sim& 4b^2 \times \frac{1}{3} \times \frac{E_n}{\omega_{Debye}} \times 
\bigg(\frac{m_n}{M_{s}}\bigg)  \nonumber \\
&\sim& 4b^2 \times \frac{1}{3} \times \frac{3meV}{10meV} \times \bigg(\frac{1}{2A}\bigg), \nonumber 
\end{eqnarray}
\noindent where $A$ is the atomic number.
As the intrinsic nuclear coherent scattering length
($b_{c}$ = 5.78 fm for the $^{16}$O nucleus)\cite{Sears}
is comparable 
to the magnetic scattering length ($r_{0}$ = 5.4 fm), the ratio of the
neutron-phonon cross section to the neutron-magnon scattering cross section
would naively be on the order of 1/A = 1/16, assuming similar kinematics. 
The results of numerical integrations
are presented in Fig.\ref{figu_total}, which show that the 
neutron-phonon scattering
is overall weaker than the neutron-magnon scattering by about one 
order of magnitude for incident neutrons with energies less than 10~meV. 
This is consistent with the above analysis. 

\section{Comparisons of Performance between S-O$_2$ and S-D$_2$}

\subsection{UCN production}

\begin{figure}[t]
\begin{center}
\includegraphics[width=3.6 in]{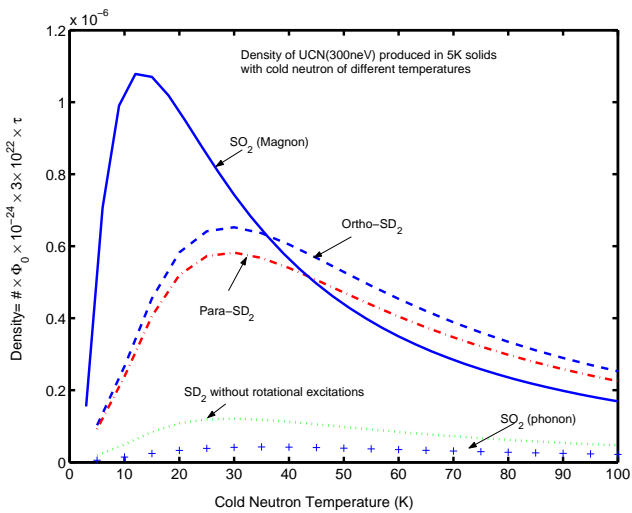}
\end{center}
\caption{UCN density vs. energy of the incident cold neutron flux characterized
by the temperature a Maxwell-Boltzmann spectrum. The UCN production  
is optimum with a flux of 12~K cold neutrons in S-O$_2$, whereas it is 
optimum with a flux of 30~K cold neutrons in S-D$_2$.} 
\label{figu_M-B}
\end{figure}

For the purpose of constructing a practical UCN source, 
we are interested not in the total cross section, but in the scattering
events that cover the UCN phase space. To estimate the UCN production
rate given a flux of incident neutrons with a broad spectrum, we carry out an
integration of the differential cross section weighted by the cold neutron
spectrum, i.e.,
\begin{equation}
P_{ucn}= n \int \bigg(\frac{d\sigma(E\rightarrow E_{ucn})}{dE}E_{ucn}\bigg) \Phi(E) dE,
\end{equation}
\noindent where $n$ is the density of the scatterers. 
We assume a Maxwell-Boltzmann energy distribution for the incident neutrons.
The results are plotted in Fig.\ref{figu_M-B} as a function of the
incident cold neutron temperature.
With a typical flux of 30~K cold neutrons on solid oxygen, 
we predict a UCN production rate of
$2.2\times 10^{-8} \Phi_0$, where $\Phi_0$ is the total incident cold neutron
flux.  As shown in the plot, the optimum cold neutron temperature for 
UCN production in solid oxygen is about 12~K, which is roughly half 
the optimum cold neutron temperature for a S-D$_2$ source. 
This gives a UCN production rate of 
$P_{ucn}= 3.5 \times 10^{-8}\Phi_{0}$, a 60\%
improvement over the 30~K cold neutrons.
For comparisons, the UCN production rates in S-D$_2$ are
$1.95 \times 10^{-8} \Phi_{0}$ (in pure ortho S-D$_2$) and
$1.74 \times 10^{-8} \Phi_{0}$ (in pure para S-D$_2$) with
30~K cold neutrons.

The maximum achievable UCN density is the production rate times the lifetime
$\tau$, i.e.,
\begin{equation}
\rho_{ucn} = P_{ucn}\times \tau.
\end{equation}
For S-O$_2$ at 2 K, if there are no loss mechanisms other than nuclear 
absorption
and magnon upscattering, $\tau = 375$ ms. By comparison, the longest
lifetimes achievable in S-D$_2$ are approximately 40 ms,\cite{Morris02} with the relative reduction due primarily
to the larger nuclear absorption cross-section, hydrogen impurities and
finite concentration of para-molecules. As a result, 
an improvement in UCN density of about an order of magnitude is expected in
solid oxygen relative to solid deuterium.

Finally, the maximum achievable UCN flux is also determined by the fiducial
source volume. For a solid deuterium source, the fiducial volume is limited
by the incoherent elastic scattering length of 8 cm.\cite{Morris02} In solid oxygen, by
contrast, the incoherent elastic scattering cross-section vanishes, so that
the elastic mean free path is in principle infinite. Thus the fiducial volume
of the solid oxygen source is expected to be limited solely by the
neutron absorption length of 380 cm.
Therefore, an improvement in source volume by a factor of up to 
\begin{equation}
\left(\frac{380}{8}\right)^d \sim  50^d \mbox{\hspace{.5in} d=1, 2, 3}
\end{equation}
\noindent over solid deuterium is theoretically
possible, though this will be limited by the maximum practical size
of a solid cryogenic source. Here, $d$ is the number of dimensions in which
the source volume can expand.
The tendency of micro-crystals to form in S-O$_2$,~\cite{Giauque29,Fage69} however,
may limit the effectiveness of a large UCN source. 

\subsection{UCN loss}

\subsubsection{Thermal Upscattering}

The 2-D symmetry of the AFM magnons in S-O$_2$ results in 
a fundamentally different dispersion curve from that of phonons.
As it turns out, it takes a finite minimum energy $E_0$ of about 8 K to
excite a magnon. For a solid oxygen sample of volume $V$ with a 
temperature below 8K,
the total number of magnons is given by 
\begin{eqnarray}
N = \sum_{i} n_i 
&\cong& \frac{V}{(2\pi)^3} \int_{E_0}^\infty Z(\epsilon)d\epsilon \bigg( \frac{1}{e^{\epsilon/k_{B}T}-1} \bigg) \nonumber \\ 
&\propto& \hspace{0.1 in} e^{-8K/T}. 
\label{T_suppress}
\end{eqnarray}
\noindent That is, the number is exponentially suppressed as the solid cools.\footnote{
In the above estimation, we approximate the density of states, $Z(\epsilon)$, to be roughly constant above $E_0$.}
For temperatures lower than 8 K, 
there are no low energy magnon levels available to be thermally populated, 
and only a small population of magnons with
energy above E$_0$ exists. This results in a partial freeze-out of 
the UCN upscattering mechanism.
The resulting exponential factor, $e^{-8K/T}$, of the magnon annihilation 
process results in a stronger temperature quenching effect on UCN upscattering
than that of the typical phonon absorption mechanism.
Note that the loss rate due to magnon upscattering
is reduced to levels comparable to the nuclear absorption rate
when the solid is cooled down to 2.1 K (see Fig.~\ref{figu_total}).
This sets the operational temperature of a solid oxygen UCN source, as
further cooling should increase the lifetime by no more than a factor of two.

\subsubsection{Para-molecule Upscattering}

Natural oxygen is over 99\% $^{16}$O, which has a nuclear spin of zero.  
The molecular rotational spectrum has only one set of allowed
states, and therefore $^{16}$O$_2$ molecules can not be left in long-lived rotational
states at low temperature. The interaction of UCN with the 
rotational states of D$_2$, however, is a major source of loss in S-D$_2$ 
sources, and must be controlled by minimizing the para-D$_2$ concentration.\cite{Liu00,Liu03}
UCN in S-O$_2$ are exempt from such a problem.

\subsubsection{Losses specific to S-O$_2$}

While nuclear absorption and magnon upscattering losses are theoretically
well understood, this is not the case for the effects of
polycrystal formation and various lattice defects.
Previous experience with solid deuterium indicates that
finite crystal effects do not
introduce additional measurable scatterings, however,
this statement might not apply to solid oxygen
because of its very different thermal properties.
This can only be answered by experiments.

In high radiation fields typical for neutron sources,
ozone formation~\cite{Bara99} in oxygen solid may be a source of UCN loss, as
the increase in mass fluctuations produces
additional incoherent scatterings.\footnote{Ozone may also be a safety concern,
although the risk is
greatly reduced in the proposed
experiment using a spallation neutron source which has a
small radiation background.}
We also expect more $\gamma$-ray heating in oxygen because of its higher Z
relative to solid deuterium. This, together with a lower thermal
conductivity,~\cite{Mucha93} may increase the demand on
cryogenic engineering to maintain the solid oxygen at 2 K.
We have also estimated the effect of additional hyperfine
transition energy in the S-O$_2$ crystal~\cite{hyperfine} on UCN loss.
While we expect no significant rate of UCN upscattering from this mechanism,
definitive measurements of the UCN lifetime in solid oxygen are required
to place quantitative limits on this and other loss mechanisms.

\section{Conclusions}

We have reviewed the potential of solid oxygen as a UCN source, 
by analyzing inelastic neutron-magnon scattering. 
Oxygen appears to be a promising
candidate, with strengths and weaknesses which  
complement those of solid deuterium.

The physics of UCN production in solid oxygen,
involving magnon (spin wave)
exchanges, is fundamentally different from the
well-known phonon mechanism in solid deuterium.
In solid oxygen, 
the neutrons are down-scattered through coupling to the
magnons in the anti-ferromagnetic $\alpha$-phase of the solid, the strength of which
is comparable to the nuclear scattering (phonon) in solid deuterium. This mechanism
has several advantages. For one,
the absence of magnon states at low energy implies
a greatly reduced UCN upscattering probability, relative to solid deuterium.
Second, the smaller nuclear absorption cross-section of
oxygen leads to a much longer UCN lifetime in this material.
Also, the UCN loss due to the absorption of the excess energy from
para-molecules (a major source of loss in solid deuterium) is absent in oxygen.

Preliminary calculations for solid oxygen indicate that the UCN density achievable should be about an order of magnitude greater than for solid deuterium.
A factor of
10 gain in source volume should be readily feasible
by taking advantage of the smaller neutron absorption probability and
the infinite elastic scattering mean free path of
UCN in solid oxygen.
Together, these properties of solid oxygen suggest that a hundredfold increase
in the UCN flux relative to what is available from solid deuterium should be
possible, given the same incident neutron flux.
The optimal operational temperature
is 2 K, somewhat lower than that of solid deuterium (5 K).


\begin{thebibliography}{99}

\bibitem{Gluck95} F. Gl\"uck, I. Jo$\acute{o}$ and J. Lasts, Nuc. Phys. A {\bf 593}i
125 (1995).

\bibitem{UCNA} UCN collaboration, High precision measurement of the neutron spin-electron correlation
asymmetry in neutron $\beta$-decay using ultra-cold neutrons.
(http://lanldb1.lanl.gov/UCN/ucnbetaasymmetrylb.nsf)

\bibitem{Lamore} I.B. Khriplovich and S.K. Lamoreaux, {\it CP Violation Without Strangeness: Electrical
Dipole Moments of Particles, Atoms and Molecules} (Springer, New York, 1997).

\bibitem{Ramsey} N.F. Ramsey, {\it In Discovery of Weak Neutral Currents: The Weak Interaction Before
and After}, eds. A.K. Mann, D.B. Cline (AIP, New York 1993);
N.F. Ramsey, Phys. Rev. {\bf 109}, 222 (1958).

\bibitem{Yuri} Y. Kamyshkov, Surv. High Energy Phys. {\bf 13}, No. 1-3, 2 (1998).

\bibitem{Golub75} R. Golub and J.M. Pendlebury, Phys. Lett., {\bf 53A},
133 (1975).

\bibitem{Ageron78}P. Ageron, W. Mampe, R. Golub and J. M. Pendlebury,
Phy. Lett. {\bf 66A} 469 (1978).

\bibitem{Morris02} C. Morris {\it et al.}, Phy. Rev. Lett. {\bf 89}, 272501 (2002).

\bibitem{solid_oxygen} P. W. Stephens and C. F. Majkrzak, Phys. Rev. B {\bf
33},
1 (1986).

\bibitem{solid_oxygen2} I. N. Krupskii, A. I. Prokhvatilov, Yu. A. Freiman
and
A. I. Frenburg, Fiz. Nizk. Temp. {\bf 5}, 271 (1979) [Sov. J. Low Temp.
Phys.
{\bf 5}, 130 (1979)].

\bibitem{raman} A. Compaan and A. Wagoner, Am. J. Phys., {\bf 62} (7),
639 (1994).

\bibitem{Lovesey84} S. W. Lovesey, {\it Theory of neutron scattering from
condensed
matter} (Clarendon press, Oxford, 1984), Vol II, ch. 4.

\bibitem{thesis} C.-Y. Liu, doctoral dissertation, physics, Princeton University, 2002 (unpublished).

\bibitem{Kleiner55} W. H. Kleiner, Phy. Rev. {\bf 97}, No. 2, 411 (1955).

\bibitem{Temme} N.M. Temme, {\it Special functions -- An introduction to the
classical functions of mathematical physics} (John Wiley \& Sons, New York,
1996), 399.

\bibitem{LEAPR} R. MacFarlane,
Technical Report No. LA-12639-MS (ENDF-356), Los Alamos National Laboratory,
Los Alamos, NM (March, 1994).

\bibitem{Sears} V. F. Sears, {\it Neutron optics: an introduction to the theory
of neutron optical phenomena and their applications} (Oxford University Press,
New York, 1989).

\bibitem{Giauque29} Giauque and Johnston, J. Am. Chem. Soc. {\bf 51}, 2300 (1929).

\bibitem{Fage69} Fagerstroem and Hallet, JLTP {\bf 1}, 3 (1969).

\bibitem{Liu00} C.-Y. Liu, A. R. Young and S. K. Lamoreaux, Phys. Rev. B
{\bf 62}, R3581 (2000).

\bibitem{Liu03} C.-Y. Liu {\it et al.}, Nucl. Instr. Meth. A {\bf 508}, 257 (2003). 

\bibitem{Bara99} Baragiola, {\it et al.}, Nucl. Instr. Meth. {\bf B} {\bf 157},
233 (1999).

\bibitem{Mucha93} A. Jezowski, P. Stachowiak, V. V. Sumarokov, and J. Mucha, Phys. Rev. Lett.
{\bf 71}, 97 (1993).

\bibitem{hyperfine} E. J. Wachtel and R. G. Wheeler, J. Appl. Phys. {\bf 42},
1581 (1971).

\end{thebibliography}
\end{document}